**Research Article**

Raquel Fernández de Cabo*, Alejandro Sánchez-Sánchez, Yijun Yang, Daniele Melati, Carlos Alonso-Ramos, Aitor V. Velasco and David González-Andrade

# Broadband mode exchanger based on subwavelength Y-junctions

**Abstract:** Multimode silicon photonics, leveraging mode-division multiplexing technologies, offers significant potential to increase capacity of large-scale multiprocessing systems for on-chip optical interconnects. These technologies have implications not only for telecom and datacom applications, but also for cutting-edge fields such as quantum and nonlinear photonics. Thus, the development of compact, low-loss and low-crosstalk multimode devices, in particular mode exchangers, is crucial for effective on-chip mode manipulation. This work introduces a novel mode exchanger that exploits the properties of subwavelength grating metamaterials and symmetric Y-junctions, achieving low losses and crosstalk over a broad bandwidth and a compact size of only 6.5 μm × 2.6 μm. The integration of SWG nanostructures in our design enables precise control of mode exchange through different propagation constants in the arms and metamaterial, and takes advantage of dispersion engineering to broaden the operating bandwidth. Experimental characterization demonstrates, to the best of our knowledge, the broadest operational bandwidth covering from 1420 nm to 1620 nm, with measured losses as low as 0.5 dB and extinction ratios higher than 10 dB. Enhanced performance is achieved within a 149 nm bandwidth (1471–1620 nm), showing measured losses below 0.4 dB and extinction ratios greater than 18 dB.

## 1 Introduction

Silicon photonics has become a fundamental technology in modern optical communication systems due to its cost-effective, high-density, and monolithic integration properties [1, 2]. Over the last decade, the significant growth in data traffic has placed ever-increasing demands on transmission capacity. The exponential growth of global Internet traffic and the rise of bandwidth-hungry applications accentuate the necessity for the telecommunications industry to move towards next-generation optical systems [3, 4]. To address this challenge, multiplexing techniques are decisive to expand the data transmission capacity of optical networks. The dominant optical transceivers in silicon photonics rely on wavelength division multiplexing (WDM) for distinct data channels [5]. However, attention in the field has shifted towards mode-division multiplexing (MDM) systems [6]. MDM techniques provide improved data rates by simultaneously transmitting several signals in optical modes of different orders within a single waveguide [7, 8]. Recent studies have also explored innovative strategies that combine WDM and MDM systems for on-chip and off-chip high-bandwidth-density interconnects [9]. Moreover, multimode silicon photonics involving higher-order modes proves to be a useful approach in emerging fields such as quantum information processing [10, 11] and on-chip nonlinear photonics [12–14].

Mode converters are fundamental components in multi-mode photonic systems, as they transform a given spatial mode into any other desired mode and vice versa [15–28]. Broadband, low-loss, and low-crosstalk are critical specifications for the overall high performance of multimode systems. Note that mode conversion structures can be categorized into mode multiplexers and mode exchangers. That is, mode exchangers transform the mode order within a single channel (i.e. a single input and a single output), whereas mode multiplexers combine modal transformations with merging multiple input channels into a single output waveguide. Many mode exchangers and multiplexers employing different structures have been proposed and, according to their respective operation principle, can be categorized into four main types [3]: phase matching [15–17], beam shaping [18–20], constructive interference of coherent scattering [21, 22] and induced gradient phase [23–25]. There are many mode exchange approaches that rely on the phase-matching condition, such as adiabatic tapers [16], asymmetric tapers [15], and chirped gratings [17]. Beam shaping mode exchange designs have been demonstrated by controlling mode evolution within a multi-mode bus waveguide that branches into separated waveguides [18, 19], or using asymmetric graded-index photonic crystals [20]. Mode exchangers based on these abovementioned principles of operation have achieved low-loss, low-crosstalk, and fabrication-tolerant performance, but suffer from narrow operational bandwidths or large footprints. However, mode exchangers relying on induced gradient phase techniques are ultra-compact, broadband, and fabrication-tolerant while maintaining low-loss and low-crosstalk. These devices can be implemented through metasurfaces [23] or metamaterials [24-28] to assist mode conversion in dielectric waveguides.

*Corresponding author: Raquel Fernández de Cabo, Instituto de Óptica, Consejo Superior de Investigaciones Científicas (CSIC), 28006 Madrid, Spain; r.fernandez@csic.es; https://orcid.org/0000-0003-4134-8735
Alejandro Sánchez-Sánchez: Telecommunication Research Institute (TELMA), Universidad de Málaga, CEI Andalucía TECH, E.T.S.I. Telecomunicación, 29010 Málaga, Spain; as.sanchez@uma.es; https://orcid.org/0009-0009-9196-2750
Yijun Yang: Centre de Nanosciences et de Nanotechnologies, CNRS, Université Paris-Saclay, 91120 Palaiseau, France; yijun.yang@c2n.upsaclay.fr; https://orcid.org/0009-0007-8215-532X
Daniele Melati: Centre de Nanosciences et de Nanotechnologies, CNRS, Université Paris-Saclay, 91120 Palaiseau, France; daniele.melati@universite-paris-saclay.fr; https://orcid.org/0000-0002-3427-0186
Carlos Alonso-Ramos: Centre de Nanosciences et de Nanotechnologies, CNRS, Université Paris-Saclay, 91120 Palaiseau, France; carlos.ramos@universite-paris-saclay.fr; https://orcid.org/0000-0002-4445-5651
Aitor V. Velasco: Instituto de Óptica, Consejo Superior de Investigaciones Científicas (CSIC), 28006 Madrid, Spain; a.villafranca@csic.es; https://orcid.org/0000-0001-7729-8595
David González-Andrade: Centre de Nanosciences et de Nanotechnologies, CNRS, Université Paris-Saclay, 91120 Palaiseau, France; david.gonzalez-andrade@c2n.upsaclay.fr; https://orcid.org/0000-0003-4402-877X

Subwavelength grating (SWG) metamaterials are artificial composite materials nanostructured with a periodic pattern where the grating period (Λ) is less than half the effective wavelength of the guided mode. No diffraction occurs at the subwavelength scale, and light propagates through the metamaterial as in a homogeneous medium with effective optical properties defined by the constituent materials and the grating geometry [29, 30]. Mode converters, in particular mode exchangers, incorporating SWG metamaterials have been successfully proven, showcasing good performance with robust fabrication tolerances and compact footprints [22,24-28]. Although multiple devices have been reported to perform mode exchange between the fundamental and first-order modes, only a few designs performing conversion into higher-order modes have been demonstrated [25], which is critical for advancing the aforementioned emerging applications [10-14]. Therefore, higher-order mode exchangers with high extinction ratios (ERs), low insertion losses (Ils) as well as compact footprints over ultra-broad wavelength ranges are still sought.

In this work, we introduce an innovative mode exchanger architecture based on two symmetric Y-junctions in back-to-back configuration connected through SWG metamaterials. The geometrical parameters of the design are optimized through the Powell's algorithm, maximizing the modal field conversion between the fundamental ($TE_0$) and second-order ($TE_2$) transverse-electric modes with a compact size of only 6.5 μm. In three-dimensional finite-difference time-domain (FDTD) simulations, the optimized device yields IL below 0.9 dB and ER exceeding 10 dB across an extensive bandwidth of 400 nm (1300 nm to 1700 nm). Within a 150 nm wavelength range (1430–1580 nm), the ER exceeds 20 dB while IL remains below 0.3 dB. To the best of our knowledge, our device reports the broadest measured operational bandwidth with IL < 0.5 dB and ER >10 dB across a 200 nm wavelength range that covers from 1420 nm to 1620 nm. Furthermore, experimental measurements demonstrate efficient mode exchange ($TE_0$-$TE_2$), showing IL as low as 0.4 dB and ER exceed 18 dB across a remarkable bandwidth of 149 nm, extending from 1471 nm to 1620 nm.

## 2 Device design

Our device architecture, depicted in Figure 1, consists of two symmetric Y-junctions arranged in back-to-back configuration and surrounded by SWG metamaterials. The geometry comprises a multimode input waveguide of width $W_S$, ensuring $TE_2$ mode support in the target wavelength range. This input waveguide is divided into two tapered waveguides of length $L_T$, which constitute the arms of the left symmetric Y-junction. The initial separation between these arms is constrained to the considered minimum feature size (MFS) and increases to a final separation $S$, while the width of these arms is reduced to the MFS. The width of the SWG segments increases along with the Y-junction separation, starting from $W_0$ = $W_S$ + 2MFS and reaching $W_F$. The length of the SWG silicon segments also exhibits a linear variation, from $a_0$ to $a_F$, for a constant Λ. The geometry is then completed with an output multimode waveguide identical to the input one, where the arms of the right Y-junction converge.

Efficient mode conversion requires matching the mode profiles of the input and output waveguides to ensure optimal coupling between the desired modes with minimal losses. When the $TE_0$ mode is excited in the input waveguide, its modal field profile evolves as it propagates through the device, ultimately being fully converted into $TE_2$ mode at the output waveguide (see section 2.1 below for details). The visual representation of the transformation of the field profile during the $TE_0$-$TE_2$ mode conversion is clearly illustrated in Figure 2(c). It is important to note that the device is reciprocal, i.e., the device also converts an incident $TE_2$ mode into the output $TE_0$ mode. Furthermore, when operating for $TE_1$ mode, the device allows it to pass through unchanged, demonstrating that the SWG structure does not affect its propagation.

### 2.1 Principle of operation

Some studies have addressed the inherent optical mode conversion characteristics of symmetric Y–junctions [31-33]. A symmetric Y–junction is composed of an input stem waveguide that branches into two output arm waveguides of equal width. The input section of a conventional symmetric Y–junction is set up as a three-layer symmetric waveguide, i.e., three dielectric layers with different refractive indices disposed as cladding-core-cladding. Here, the layers refer to the horizontal arrangement ($y$ direction) of the dielectric regions forming the waveguide structure. Likewise, the output arm waveguides of a conventional symmetric Y–junction exhibit a five-layer symmetric structure: cladding-core-cladding-core-cladding. The intrinsic symmetry of these devices restricts mode conversions exclusively to even-to-even or odd-to-odd modes. The amount of power conversion between modes can be determined through the overlap integral computed at the interface between stem and arms waveguides, involving incident and transmitted fields. The incident field ($E^i_z$) corresponds to that present in a typical three-layer slab waveguide. The field transmitted to the output branching arms waveguides can be approximated by [31-33]:

$$E_z^t(y) = B(d)E_z(y,d)e^{\left(-j\cos\alpha \int_0^x \beta(x')dx'\right)}e^{(-j\beta|y|\sin\alpha)}$$

where the tilt of the arm waveguides is expressed by $\alpha$ = atan($S/2L_T$), and the separation between the branching waveguides (2$d$) is a function of $x$: $d = x \tan(\alpha)$. The electric field $E_z(y,d)$ is the corresponding guided-TE mode in a symmetric five-layer waveguide and $B(d)$ is a normalizing factor. The quantification of transmitted power between the input mode of order $m$ into the output mode of order $m'$ is contingent upon the branching angle and the difference between its propagation constant ($\beta$) in the three-layer ($\beta^i_m$) and five-layer ($\beta^t_{m'}$) guided wave mode. This can be expressed by the transmitted coupling coefficient:

$$c^t_{m,m'} = \frac{1}{\omega\mu_0} \frac{\beta^i_m \beta^t_{m'}}{\beta^i_m + \beta^t_{m'}} \int_{-\infty}^{\infty} E^i_{z,m}(E^t_{z,m'})^* dy$$

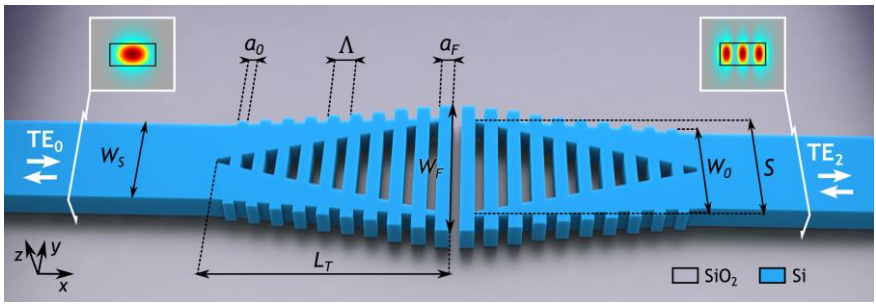

**Fig. 1:** Three-dimensional schematic of the proposed SWG mode exchanger. Insets depict the distribution of the electric field magnitude ($|E_y|$) in the cross-sectional view of the input/output waveguides.

This coefficient is essential for predicting the performance of mode conversion in the stem-to-arms interface of symmetric Y-junctions. It is important to note that the propagation constants ($\beta^t_{m'}$) of the transmitted modes, governed by the eigenvalue equations for the five-layer waveguide structure, change as the separation between the arm waveguides increases.

Within the proposed structure, the symmetric Y-junction initiates modal conversion, and the propagating modes within the five-layer structure experience a phase difference. This phase difference arises from the respective phases of the propagating modes through the arms and the central SWG metamaterial, thereby achieving complete mode exchange between the TE$_0$ and TE$_2$ modes. Furthermore, the strategic integration of SWG metamaterials allows us not only to induce the proper phase difference but also to achieve a wavelength-independent response. Previous studies have demonstrated that SWG metamaterials can be engineered for wavelength insensitivity by exploiting dispersion engineering [34,35]. In our design, the TE$_0$ and TE$_2$ modes propagating through the five-layer structure comprising SWG metamaterials experience a phase difference ($\Delta\phi$) that can be expressed in terms of the cumulative phase shift for each period:

$$\Delta\phi = \sum_{k=1}^{N_{periods}} \left(\beta^k_0 - \beta^k_2\right)\Lambda$$

By properly selecting the geometrical parameters of the grating, the difference between the propagation constants of the TE$_0$ ($\beta_0$) and TE$_2$ ($\beta_2$) modes can be made independent of wavelength. Therefore, the incorporation of SWG metamaterials between the arms of the Y-junction in our device facilitates precise control over mode exchange, enhancing the device's operational response across a broad range of wavelengths. However, the integration of these periodic nanostructures comes at the cost of a simultaneous increase in design complexity. Additionally, some geometrical parameters of the device vary along the propagation direction. For instance, the width of the arm waveguides is progressively reduced and the width and length of the SWG segments change. Consequently, the optimization of this complex geometry results in a significant design challenge due to the multiple parameters that drastically affect the device's response. To address this challenge, we use an optimization algorithm that maximizes the efficiency of the mode conversion by iteratively refining the design parameters to achieve optimal performance.

## 2.2 Optimization process

The proposed mode exchanger is optimized through an inverse design approach using Powell's algorithm [36]. This algorithm involves an iterative optimization process that systematically explores conjugate directions within a given parameter space to maximize the target figure of merit (FOM). The parameter space involved in this approach contains the critical geometric parameters of both Y-junctions and subwavelength gratings. The algorithm is initialized with a parameter set and sequentially performs conjugate directional and line searches, thus allowing adaptive updates of the parameter values. This method efficiently navigates high-dimensional spaces, making it suitable for complex multi-parameter optimizations. Additionally, the algorithm considers parameter bounds, restricting geometric parameter values to align with practical design constraints, such as fabrication MFS. Specifically, our figure of merit represents the modal field forward transmission ($T_{fwd}$) from the incident TE$_0$ mode into the targeted output TE$_2$ mode across the operational wavelength range, obtained through 3D-FDTD simulations. In an ideal scenario, the target transmission ($T_{target}$) value should be equal to one, indicating lossless modal conversion between the specified input and output modes. Therefore, the FOM in this context is a numerical value that considers the transmission values and their integrals, normalized by the wavelength range:

$$FOM = \int_{\lambda_1}^{\lambda_2} \frac{|T_{target}|}{\lambda} d\lambda - \int_{\lambda_1}^{\lambda_2} \frac{|T_{fwd} - T_{target}|}{\lambda} d\lambda$$

The proposed mode exchanger is developed for a silicon-on-insulator (SOI) platform with a 220-nm-thick silicon core layer. The MFS of the design is fixed at 60 nm in order to be compatible with e-beam fabrication process. The following geometric parameters, depicted in Figure 1, are held constant during the optimization process. For the multimode input and output waveguides, $W_S$ is set at 1.4 µm, ensuring TE$_2$ mode support meanwhile suppressing higher-order modes for wavelengths between 1300 nm and 1700 nm. For the SWG segments, $W_0$ is fixed at 1.52 µm, to ensure the MFS restriction, and $\Lambda$ = 200 nm is selected to avoid radiation and Bragg regimes. On the other hand, the geometric parameters subject to optimization include the duty cycles characterizing the narrower ($DC_0 = a_0/\Lambda$) and wider ($DC_F = a_F/\Lambda$) segments of the SWG, as well as $W_F$, $S$, and $L_T$. Table 1 presents the parameter values (see Figure 1) used for algorithm initialization, including the paired boundary values that define the minimum and maximum permissible ranges for each parameter. The table further shows the refined parameter values achieved through the Powell algorithm after 100 iterations, highlighting the compact total length of only 6.5 µm. The number of iterations was chosen to ensure a comprehensive exploration of the design space while maintaining computational efficiency. This choice was further validated by the high FOM attained compared to the state of the art and by its stability within the last 15 iterations.

Figure 2(a) shows the FOM over iterations during optimization, with deep drops resulting from specific geometric parameter adjustments that minimize conversion efficiency. Figure 2(b) presents the insertion loss and extinction ratio for the resulting design which incorporates the final values of the geometric parameters (presented in Table 1). The inset of Figure 2(b) provides a zoomed-in view of the device losses. Figure

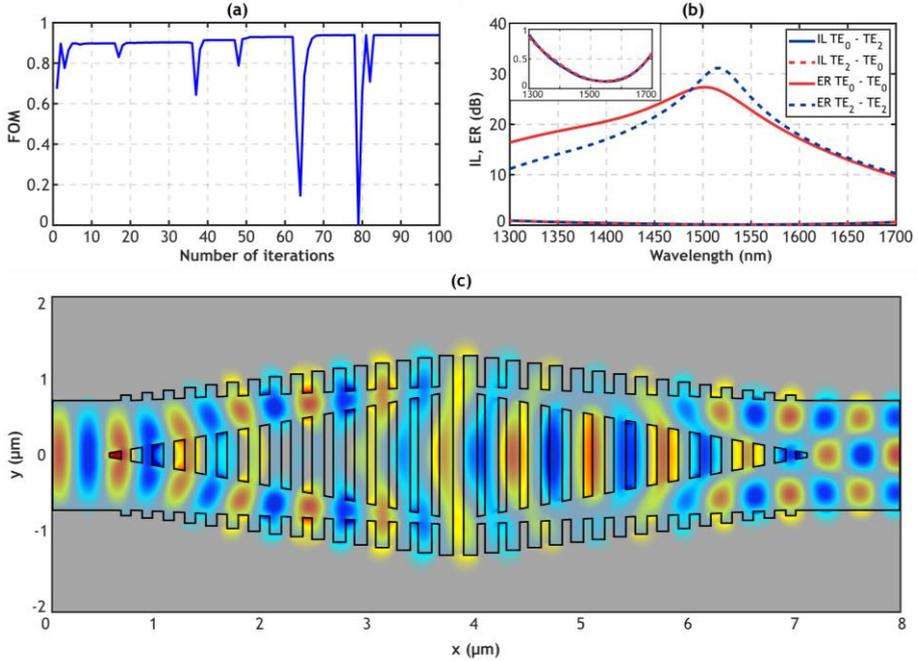

**Fig. 2:** (a) Evolution of the figure of merit for each iteration of the optimization process. (b) Insertion loss and extinction ratio of the final optimized design for $TE_0$ input, with an inset of a zoomed-in view of the device losses. (c) Real part of the electric field ($E_y$) propagating along the device obtained by 3D-FDTD simulation.

2(c) additionally shows the propagation of the real part of the electric field ($E_y$) along the device. The insertion loss and extinction ratio are defined as IL = $-10 \cdot \log_{10}(T_{0,2})$ and ER = $10 \cdot \log_{10}(T_{0,2}/T_{0,0})$, respectively. Here $T_{0,2}$ and $T_{0,0}$ represent the transmission from an incident $TE_0$ mode into the $TE_2$ mode ($T_{0,2}$) and $TE_0$ mode ($T_{0,0}$) at the output, respectively. Our device demonstrates IL under 0.9 dB and ER over 10 dB in a 400 nm broad bandwidth from 1300 nm to 1700 nm. Moreover, in a 150 nm bandwidth covering the 1430 – 1580 nm wavelength range, ER is higher than 20 dB and IL remains below 0.3 dB.

**Tab.1:** Optimization process parameter values.

| Parameter | Initial value | Bound min | Bound max | Final value |
|---|---|---|---|---|
| $DC_0$ | 60% | 35% | 70% | 47% |
| $DC_F$ | 60% | 35% | 70% | 65% |
| $W_F$ | 2.6 μm | 2.0 μm | 3.6 μm | 2.6 μm |
| $S$ | 1.6 μm | 1.0 μm | 2.0 μm | 1.8 μm |
| $L_T$ | 3.2 μm | 1.0 μm | 4.0 μm | 3.236 μm |

## 3 Measurements

The device comprising the final parameter values (presented in Table 1) was fabricated in an SOI platform with a 220-nm-thick silicon core layer and 3-μm buried oxide. The necessary patterns were defined using electron-beam lithography (RAITH EBPG 5200) and then transferred using reactive ion etching (ICP-DRIE SPTS). Scanning electron microscope (SEM) images of the fabricated device were collected before the top cladding was deposited (Figure 3). Then, a 1.5-μm-thick layer of polymethyl methacrylate (PMMA) was applied to the sample using spin-coating.

The devices were characterized using two tunable lasers to sweep the 1420 – 1510 nm and 1510 – 1620 nm wavelength ranges. The light was coupled to the chip and collected from the chip using cleaved fibers. Optimized grating couplers for TE polarization were integrated for precise light injection and extraction from the fiber to the chip. The on-chip characterization structure consisted of a mode multiplexer-mode exchanger-mode demultiplexer configuration with three input and three output single-mode waveguides as reported in [7]. By selectively injecting light into the first, second, or third input waveguide of the mode multiplexer, the $TE_0$, $TE_1$, or $TE_2$ mode was respectively excited at the input waveguide of the mode exchanger. Subsequently, after the light passes through the mode exchanger, the output demultiplexer reciprocally extracted each mode as the fundamental mode of its corresponding output waveguide. Specifically, the $TE_0$, $TE_1$, or $TE_2$ mode proceeding from the mode exchanger was effectively coupled to the fundamental mode of the first, second, or third output waveguide, respectively. In addition, the chip also included reference waveguide structures mirroring that of the mode multiplexer-mode exchanger-mode demultiplexer but replacing the mode exchanger with a multimode waveguide of the same length ($2L_T + \Lambda - a_F$) and width ($W_S$). The response of these reference structures was integrated in IL and ER extraction.

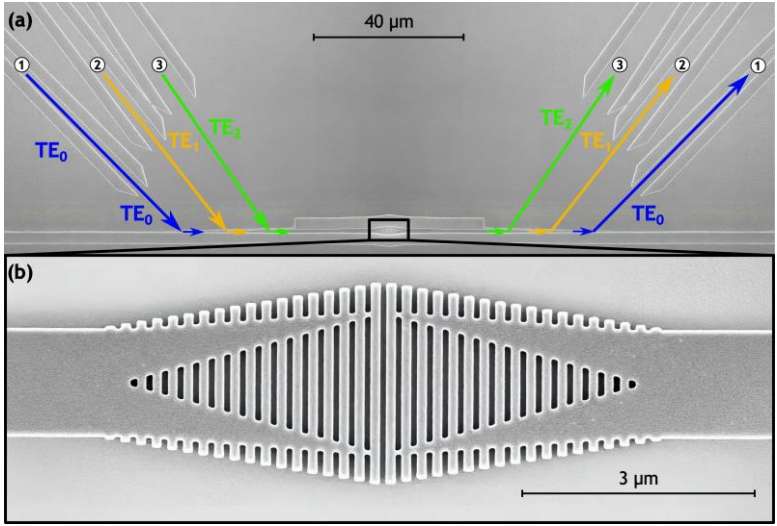

**Fig. 3:** (a) Scanning-electron-microscope images of (a) the mode multiplexer architecture used for the device characterization [7] and (b) a magnified view of the fabricated mode exchanger.

Figure 4(a), and (b) present the ILs and ERs measured when injecting the $TE_0$, $TE_1$, or $TE_2$ mode, respectively. Figure 4(a) illustrates the conversion from the input $TE_0$ mode into the output $TE_2$ mode (solid lines), demonstrating experimental IL under 0.35 dB across the measured bandwidth ranging from 1420 nm to 1620 nm. Inset providing a close-up view of the measured IL have also been included in Figure 4. The reciprocal process of $TE_2$-$TE_0$ modal conversion is also depicted in Figure 4(a) with dashed lines, exhibiting measured IL below 0.52 dB in the same 200 nm bandwidth. In both instances of mode conversion, specifically $TE_0$-$TE_2$ and $TE_2$-$TE_0$, the device exhibits IL as low as 0.4 dB and ER exceed 18 dB across a broadband of 149 nm extending from 1471 nm to 1620 nm. Additionally, ER remains above 10 dB over the whole 200 nm measured bandwidth. As shown on Figure 4(b), our proposed device does not affect the propagation of the $TE_1$ mode. When the $TE_1$ mode is injected into the mode exchanger, it efficiently propagates through the device with IL under 0.6 dB and ER above 17 dB within the measured wavelength range of 1420 – 1620 nm. These experimental results demonstrate the effectiveness of the device achieving high extinction ratios and ensuring efficient conversion with minimal losses.

## 4 Discussion and conclusions

We have proposed and experimentally demonstrated a metamaterial-engineered mode exchanger for TE polarization that operates over a broad bandwidth. The device design

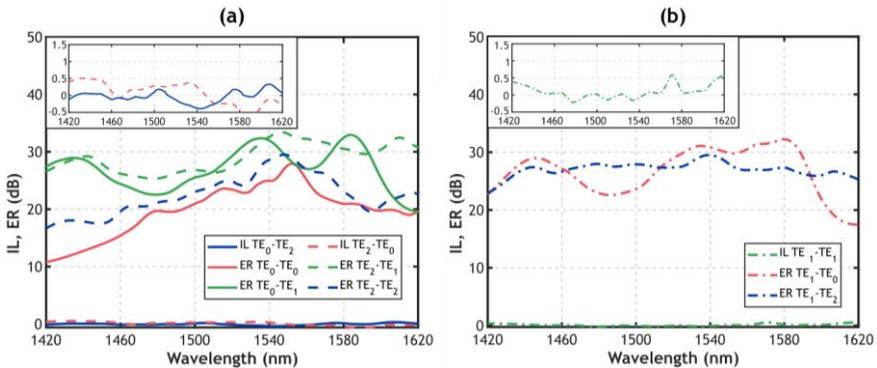

**Fig. 4:** Measured insertion loss and extinction ratio for $TE_0$ mode (red curves), $TE_1$ mode (green curves), and $TE_2$ mode (blue curves) when injecting (a) $TE_0$ mode (solid lines) and $TE_2$ mode (dashed lines) and (b) $TE_1$ mode (dash-dot lines). Insets of the measured IL have been included to improve readability.

**Tab.2:** Comparison of state-of-the-art $TE_0$-$TE_2$ mode exchangers.

| Ref. | Structure | Length (μm) | Simulation | | | Experiment | | |
|---|---|---|---|---|---|---|---|---|
| | | | IL (dB) | ER (dB) | BW (nm) | IL (dB) | ER (dB) | BW (nm) |
| [15] | Asymmetric tapers | 19.317 | <0.08* | NR | 60 (1520-1580) | | NF | |
| [23] | Shallow-etched SWG | 6.736 | <1* | >12.5* | 20 (1545-1565) | <0.5 | >10 | 20 (1545-1565) |
| [24] | SWG and asymmetric tapers | 4.9 | <0.8 | >15 | 100 (1510-1610) | <1.8 | >10 | 81 (1480-1561) |
| [25] | SWG phase-shift section | 2.9 | <0.85 | >10 | 340 (1354-1694) | <1 | >13 | 87 (1520-1607) |
| [37] | Slot waveguides | 2.4 | <0.22 | >18 | 50 (1520-1570) | <0.3 | >9 | 50 (1520-1570) |
| [38] | Substrip dielectric waveguides | 2.2 | <2* | >12* | 300 (1400-1700) | <3* | >12* | 45 (1535-1580) |
| [39] | Polygonal slot waveguides | 24 | <0.1* | >20 | 100 (1500-1600) | | NF | |
| [40] | SWG-assisted tapers | 4.98 | <0.19 | >17.36 | 100 (1500-1600) | | NF | |
| [41] | Waveguide-width-modulation | 6.5 | <0.283 | NR | 100 (1500-1600) | | NF | |
| [42] | Two-etch bragg grating | 65.3 | <3 | >21 | 24.5 (1540.5-1565)* | | NR | |
| This work | SWG symmetric Y-junctions | 6.542 | <0.9 | >10 | 400 (1300-1700) | <0.4 | >18 | 149 (1471-1620) |

Values marked with an asterisk correspond to estimations from figures. NR, not reported; NF, not fabricated; IL, insertion loss; ER, extinction ratio; BW, bandwidth.

presented here achieves efficient mode exchange, leveraging the inherent modal properties of symmetric Y–junctions combined with SWG metamaterials. Moreover, the strategic engineering of SWG nanostructures along the Y-junction structure enables the further exploitation of metamaterial dispersion, achieving experimentally remarkable high extinction ratios over a very broad bandwidth. Powell's optimization algorithm is used to maximize the figure of merit, representing the modal field transmission from $TE_0$ mode to $TE_2$ mode, by iteratively adjusting all relevant design geometric parameters. In 3D FDTD simulations, the final optimized design demonstrates IL below 0.9 dB and an ER exceeding 10 dB over a broad bandwidth covering from 1300 nm to 1700 nm. Within a 150 nm wavelength range (1430 – 1580 nm), ER exceeds 20 dB while IL remains below 0.3 dB.

Experimental measurements demonstrate the high performance of the design, showcasing effective mode conversion between fundamental and second order TE modes, with low IL and high ER. Specifically, for $TE_0$-$TE_2$ mode exchange, the IL is below 0.35 dB, and for the reciprocal process ($TE_2$-$TE_0$), it is under 0.52 dB across the entire measured bandwidth from 1420 nm to 1620 nm. In both instances of mode conversion, specifically $TE_0$-$TE_2$ and $TE_2$-$TE_0$, the device exhibits IL as low as 0.4 dB and ER exceed 18 dB across a broadband of 149 nm extending from 1471 nm to 1620 nm.

Table 2 provides a summary of the performance of the demonstrated device alongside several state-of-the-art $TE_0$-$TE_2$ mode exchangers, in order to perform a comparative analysis. Among all the reviewed state-of-the-art devices, the largest bandwidth simulated and experimentally reported is achieved using a SWG phase-shift region in ref. [25]. This device is a $TE_0$-$TE_2$ mode exchanger, in this case leveraging two symmetric SWG structures to induce controlled phase shifts within a multimode waveguide, which achieves experimental ER > 13.2 dB over a bandwidth of 87 nm (1520 – 1607 nm). In contrast, our structure, consisting of two symmetric Y-junction with SWG metamaterials, demonstrates a measured ER > 18 dB within a 149 bandwidth (1471-1620 nm). The mode exchanger proposed in ref. [37] exhibits the lowest measured losses, showing IL of less than 0.3 dB and ER greater than 9 dB in a 50 nm bandwidth. The device reported in ref. [38] exhibits the second broadest simulated bandwidth, with IL less than 2 dB and ER greater than 12 dB over a wavelength range covering 300 nm, but only proving experimental bandwidth of 45 nm. In comparison to these alternative solutions, the mode exchanger introduced in this work represents a significant improvement over current state-of-the-art devices, offering enhanced functionality in several aspects. It notably outperforms existing devices in terms of optical bandwidth, showing almost a two-fold increase over the most advanced experimental demonstrations to date. Furthermore, the device achieves highly efficient mode conversion, with an insertion loss of less than 0.4 dB and an extinction ratio exceeding 18 dB over a wavelength range of 149 nm. Notably, such performance is achieved while maintaining a compact design with an overall length of only 6.54 μm.

The presented mode exchanger stands as a pioneering solution in the field, distinguished by its notable combination of augmented bandwidth, efficiency, and compact design. These compelling attributes underline the device's considerable potential in ultrahigh-density MDM systems for advanced multimode optical communication and signal processing applications. The device is also expected to find application in emerging fields, such as in quantum information processing or on-chip nonlinear photonics. Moreover, optimizing the design of this structure could potentially enable the simultaneous excitation of $TE_0$ and $TE_2$ modes from an initial $TE_0$ mode input, which is crucial for quantum computing applications relying on transverse-mode encoding.

**Research funding:** This work has been funded by the Spanish Ministry of Science and Innovation (PID2020-115353RA-I00); the Spanish State Research Agency and the European Social Fund Plus under grant PRE2021-096954; the European Union's Horizon Europe (Marie Sklodowska-Curie grant agreement No. 101062518 and HADEA grant agreement No. 101135523). This work was done within the C2N micro nanotechnologies platforms and partly supported by the RENATECH network and the General Council of Essonne.

**Author contribution:** All authors have accepted responsibility for the entire content of this manuscript and approved its submission.

**Conflict of interest**: Authors state no conflict of interest.

**Informed consent**: Informed consent was obtained from all individuals included in this study.

**Data availability statement**: The data that support the findings of this study are available from the authors upon reasonable request.